\documentclass{ws-ijmpe}

\begin{document}



\title{Single-Event Handbury-Brown-Twiss Interferometry}

\author{CHEUK-YIN WONG}

\address{Physics Division, Oak Ridge National Laboratory,
Oak Ridge, TN 37830, U.S.A.\\
Department of Physics, University of Tennessee, Knoxville, TN
37996, U.S.A.\\
wongc@ornl.gov}

\author{WEI-NING ZHANG}

\address{
Department of Physics, Dalian University of Technology, 
Dalian, Liaoning 116024, China\\
Department of Physics, Harbin Institute of Technology, 
Harbin, Heilongjiang 150006, China\\
weiningzh@hotmail.com}

\maketitle

\begin{history}
\received{(received date)}
\revised{(revised date)}
\end{history}

\begin{abstract}
Large spatial density fluctuations in high-energy heavy-ion collisions
can come from many sources: initial transverse density fluctuations,
non-central collisions, phase transitions, surface tension, and
fragmentations.  The common presence of some of these sources in
high-energy heavy-ion collisions suggests that large scale density
fluctuations may often occur.  The detection of large
density fluctuations by single-event Hanbury-Brown-Twiss
interferometry in heavy-ion collisions will provide useful information
on density fluctuations and the dynamics of heavy-ion
collisions.
\end{abstract}

\section{Introduction}

Large spatial density fluctuations in high-energy heavy-ion collisions
can come from many sources.  Some of these sources are expected to be
quite common in high-energy heavy-ion collisions and consequently
large scale density fluctuations may often occur.  It is of great
interest to examine the sources of large density fluctuations and to
find ways for their detection so as to obtain useful information on
the collision dynamics\cite{Zha95,Zha04,Won04,Zha05,Zha06,Zha06a}.

Among the many different ways to study a quark-gluon plasma, intensity
interferometry (HBT interferometry) is one of the experimental tools
to examine the space-time density distribution of the produced
matter\cite{Won94}.  The usual application of the HBT interferometry
uses pairs of identical particles from many events (multi-event
analysis) of similar general characteristics to study the (averaged)
space-time configuration.  Because of the difference in initial
conditions in different collision events, different collisions will
lead to different density fluctuations for some degrees of freedom
while retaining similar fluctuations for some other degrees of
freedom.  An average over many events in a standard HBT analysis will
lead to a distribution with suppressed fluctuations in event-dependent
degrees of freedom but will retain those fluctuations in
event-independent degrees of freedom.  The proposed single-event HBT
analysis, if it can be carried out with sufficient accuracy, allows
the examination of the full characteristics of large scale density
fluctuations without suppression.  The observation of HBT
interferometry in atomic systems also make it interesting to examine
the interference pattern in systems with large macroscopic density
fluctuations\cite{Nar99,Gom06}.

If the statistics of identical bosons turns out to be insufficient for
single-event HBT analyses using presently available detectors, it will
be of interest to carry out few-event HBT analyses to examine 
event-independent density fluctuations. The few-event analysis will
however require a good theoretical understanding of the single-event
HBT interferometry.

With the example of a granular density distribution to indicate the
type of correlations one can get in a large scale density fluctuation,
we shall show that the correlation function of a highly fluctuating
density distribution is likewise highly fluctuating, with maxima and
minima at locations which depend on the relative coordinates of the
density centers.  These local maxima and minima arise from the
constructive and destructive interferences of identical bosons
originating from two different centers.  Their interference patterns
therefore provide useful information on highly fluctuation density
distributions in high-energy heavy-ion collisions.

\vspace*{0.2cm}
\section{Sources of Large Density Fluctuations}

In many simulations of the heavy-ion collisions on an event-by-event
basis, the initial transverse energy density is far from being
uniform.  It exhibits large transverse density fluctuations with a
large peak-to-valley ratio\cite{Dre02,Ham05,Soc04}.  A number of
transverse density `lumps' are clearly visible in the initial
transverse density distribution of the produced matter in Fig. 21 of
Ref.\ [\cite{Dre02}] or Fig. 4 of Ref.\ [\cite{Ham05}] shown in Fig.
1(a).

\vspace*{-0.9cm}
\begin{figure}[h]
\hspace*{2.8in}
\psfig{file=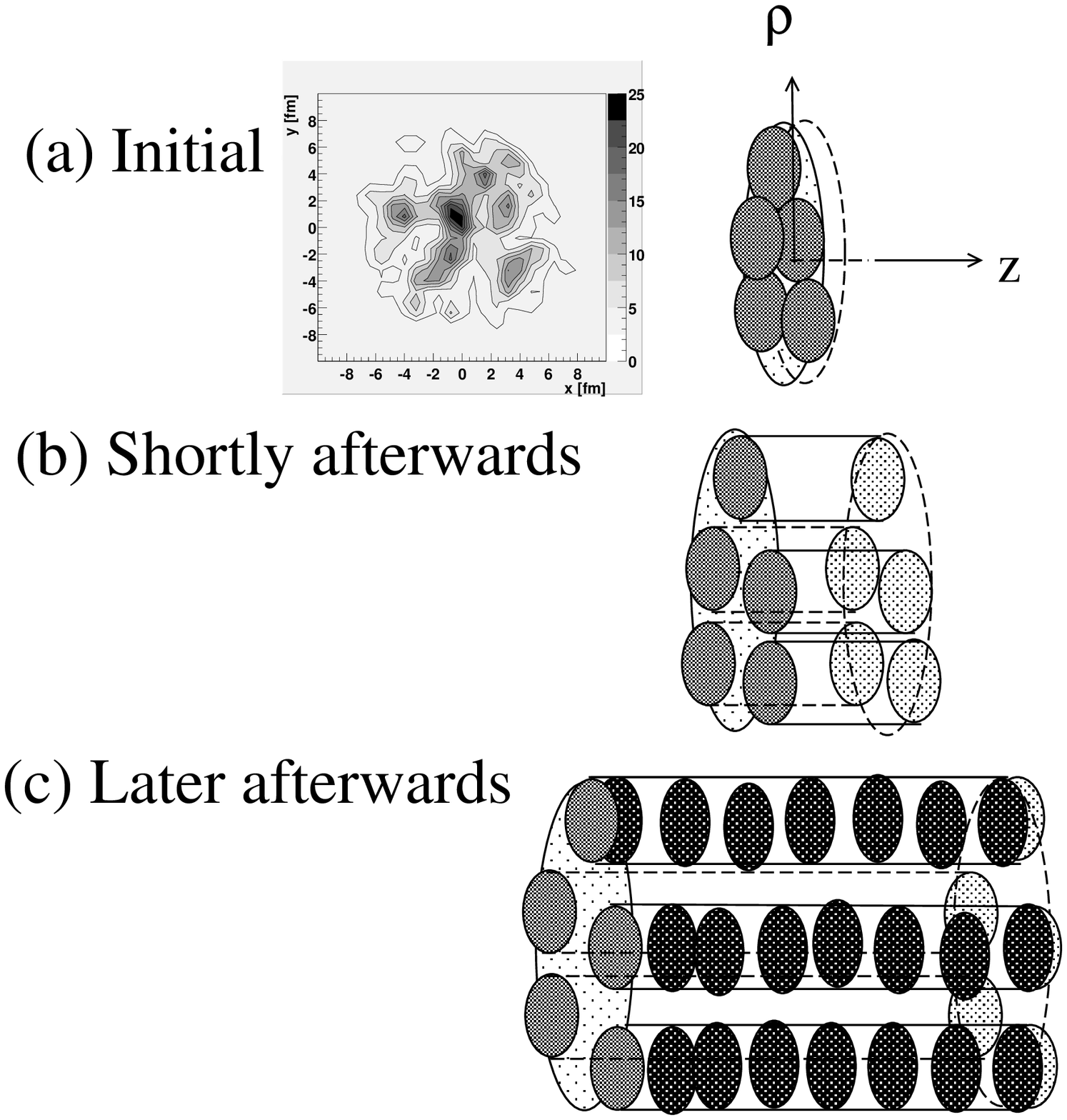,width=5cm}
\end{figure}

\vspace*{-0.7cm}
\hangafter=0
\hangindent=2.85in
\noindent Fig. 1.   The scenario in a central high-energy heavy-ion collision.

\vspace*{-6.3cm} \hangafter=-15 \hangindent=-2.3in Because of Lorentz
contraction, the longitudinal length of the produced matter is much
smaller than the transverse length.  Initially as depicted in
Fig. 1(a), density fluctuations in the transverse plane manifest as a
number of transverse lumps\cite{Dre02,Ham05}. The subsequent
longitudinal expansion proceeds much faster than the transverse
expansion, as pointed out by Landau and Belenkii\cite{Lan53}.  The
expanding matter develops into tubes [Fig.\ 1($b$)] and the density of
matter in the tube decreases as a function of the proper time.  The
sausage instability leads to the break-up of the tube of matter into
many approximately equal spherical droplets\cite{Zha06} depicted in
Fig. 1(c), as in the break-up of toroidal liquid drops\cite{Won73}.

 \hangafter=5 \hangindent=-2.3in Another initial transverse density
distribution occurs in non-central collisions of two approximately
equal size nuclei.  In these collisions, while the spectator matter
proceeds forward, the participants collide and interact.  The
pion-emitting source produced by the participants may retain part of
the initial momenta in different spatial regions.  If the produced
matter constituents retain a large fraction of the initial momenta,
they can be approximately idealized as two pion emitting clusters as
shown in Fig. 2$a$.  On the other hand, if a significant fraction of
the participant matter in the overlapping region is stopped while the
more peripheral constituents moves in their original directions, the
pion emitting source can be approximated as consisting of three
clusters as in Fig. 2$b$.  Interference of identical pions emitted
from either two or three clusters will exhibit features associated
with these spatial configurations.

\vspace*{-6.0cm}
\begin{figure}[h]
\hspace*{2.8in}
\psfig{file=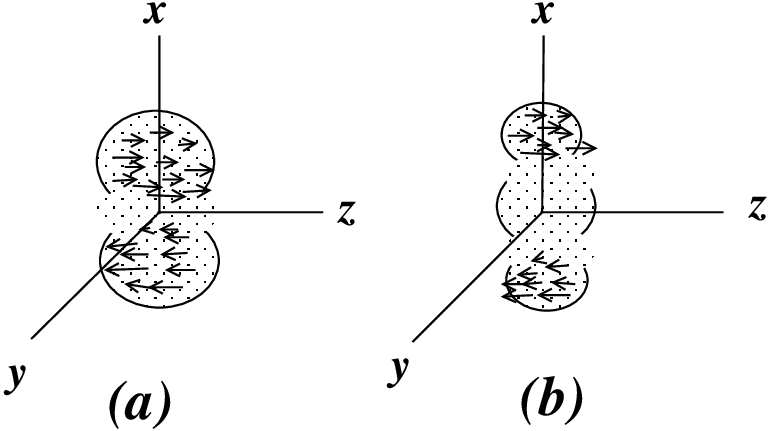,width=5cm}

\end{figure}

\vspace*{-0.6cm}
\hangafter=0
\hangindent=3.0in
\noindent Fig. 2.  Possible configurations of the pion-emitting source
after a non-central collision.

\vspace*{ -0.0cm}
\hangafter=-2 \hangindent=-2.3in 
Another source of density fluctuation arises from the occurrence of
first-order phase transition, as was first pointed out by
Witten\cite{Wit84} and examined by many
workers\cite{Zha95,Zha04,Won04,Zha05,Zha06,Pra92}.  According to such
a description, incipient bubbles of low-density matter (hadrons) are
formed in some local regions in a nearly static quark-gluon plasma, as
the local temperature decreases below the phase transition
temperature.  The coalescence of these bubbles leads to bubbles of
larger radii.  As the system cools further, the fraction of matter in
the low-density bubbles will become greater than the fraction of
high-density matter (QGP), and the high-density matter will turn into
lumps in the low-density matter background.  The density of such a
configuration is highly fluctuating.

Still another source of density fluctuation arises from the effects of
surface tension during the evolution of the strongly-interacting
matter. Because the QGP and the hadron matter are strongly interacting
dense media, a surface tension arises at a boundary due to the
presence of a strong interaction in the dense phase on one side of the
boundary and the absence (or weakening) of the strong interaction with
no density (or diminishing density) on the other side.  This imbalance
of the forces acting on different parts at the boundary leads to the
surface tension, and the detail profile of the boundary may depend
on the nature and the order of phase transition.  Due to the presence
of this surface tension, there may be the instability against surface
shape changes that favors the formation of surfaces of smallest areas
in different local regions.  Such a tendency lead to large spatial
density fluctuations.  The end product may be the multi-fragmentation
of the strongly-interacting matter.  Multi-fragmentation has been
observed in intermediate energy heavy-ion collisions\cite{Riz06}.

\section{Detection of Large Density Fluctuations by Single-Event 
HBT Interferometry}

In order to understand how the features of the correlation function
may be affected by the presence of large density fluctuations, it is
worth studying the correlation function of identical particles of
momentum $k_1$ and $k_2$ with $q=k_1-k_2$ and $P=(k_1+k_2)/2$ for an
idealized chaotic droplet sources\cite{Won04}.  There is a relation
between the two-boson correlation function and the Fourier transform
of the source density $\rho(x;k_1,k_2)$, which we assume to be
independent of $q$,
\begin{eqnarray}
C(q,P)=1
+\bigl |\int dx e^{iq\cdot x} \rho(x,P) \bigr |^2.
\end{eqnarray}
For brevity of notation, we shall leave the label $P$ implicit
and understood.

We consider a density distribution of $N$ droplets of the type
\begin{eqnarray}
\rho (x) = A\sum_{j=1}^N  \rho_j(x),
\end{eqnarray}
where $\rho_j$ is the density distribution of the $j$-th droplet, and
$A$ is a normalization constant such that the total density $\rho(x)$
is normalized to unity as $\int ~dx~ \rho(x)= 1$.  We can consider the
$j$-th droplet to have particle emission time centering at $T_j$, to
be localized initially at ${\bf R}_j$ with standard deviations
$\sigma_j$ and $\tau_j$, and to move with a velocity ${\bf v}_j$,
\begin{eqnarray}
\label{rhodis}
\rho_j(t,{\bf r}) = \frac{ e^{-({\bf r}-{\bf R}_j-{\bf v}_jt)^2/2\sigma_j^2}}
                       { (\sqrt{2\pi}\sigma_j )^3  }
                   \frac{ e^{-(t-T_j)^2/2\tau_j^2} }
                       { \sqrt{2\pi}\tau_j  }.
\end{eqnarray}
The correlation function is then
\begin{eqnarray}
C(q)=1+\left |A \sum_{j=1}^N  \exp \biggl \{
i {\bf q} \cdot{\bf R}_j-
i(q_0-{\bf q}\cdot{\bf v}_j)T_j
-\frac{\sigma_j^2{\bf q}^2}{2}
-\frac{\tau_j^2(q_0-{\bf q}\cdot{\bf v}_j)^2}{2}
 \biggr \}
\right | ^2.
\end{eqnarray}

In order to get a clear insight into the most important features of
the correlation function, we consider the simple case where the
density distributions of all droplets are the same so that
$\sigma_j=\sigma$ and $\tau_j$=$\tau$ for all $j$.  In this simple
case, $A=1/N$ and the correlation function can be easily evaluated to
give
\begin{eqnarray}
\label{eq31}
  C(q)=1+\frac {e^{-\sigma^2 {\bf q}^2}}
               {N^2}
\left [ \sum_{j=1}^Ne^{-\tau^2(q_0-{\bf q}\cdot {\bf v}_j)^2}
    + 2\sum_{\scriptstyle j,k=1 \atop j>k \scriptstyle}^N  
e^{-\frac{\tau^2}{2} [(q_0-{\bf q}\cdot {\bf v}_j)^2 + (q_0-{\bf q}\cdot {\bf v}_k)^2]}
\cos \Delta_{jk}
 \right ],
\end{eqnarray}
\begin{eqnarray}
\Delta_{jk}=-q\cdot (X_j-X_k)-{\bf q}\cdot({\bf v}_jT_j-{\bf v}_kT_k),
~~~~X_j=(T_j,{\bf R}_j).
\end{eqnarray}
Thus, the correlation function $C(q)$ has maxima at $\Delta_{jk}\sim
2n\pi$, and minima at $\Delta_{jk} \sim(2n-1)\pi$, with $n=1,2,3,...$.
Numerical examples with $N=4$ and $N=8$ static droplets are shown in
Figs. 1 and 2 of Ref. [\cite{Won04}].  The number of correlation
function maxima for 8 droplets is greater than the number of maxima
for 4 droplets.  The smaller the number of droplets, the greater will
be the magnitude of fluctuations, as can be easily deduced from Eq.\
(\ref{eq31}).  Previously, we examine the two-pion interferometry for
a granular source of quark-gluon plasma droplets and found that the
granular model of the emitting source may provide an explanation to
the RHIC HBT $R_{\rm out}/R_{\rm side}$ puzzle\cite{Zha04,Zha06}.  We also
investigate two-pion Bose-Einstein correlations of many droplets in
single-event measurements.  We find that the distribution of the
fluctuation between correlation functions of the single- and
mixed-events provide useful signals to detect the granular structure
of the source\cite{Zha05}.

\vspace*{-0.4cm}
\begin{figure}[h]
\hspace*{1.4cm}
\psfig{file=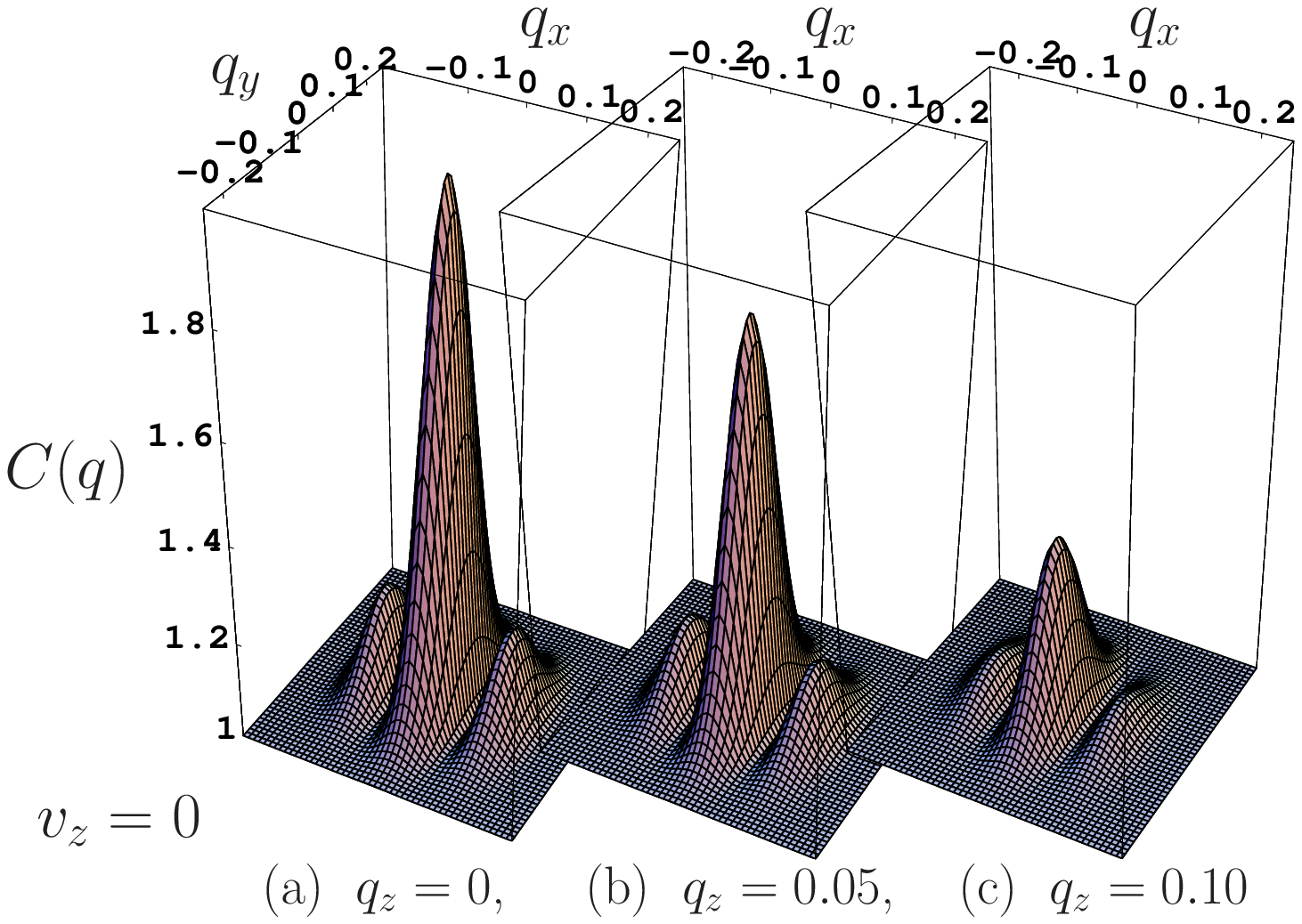,width=10cm}
\end{figure}

\vspace*{-0.4cm}
\hangafter=0
\hangindent=2.0cm
\noindent Fig. 3.   The correlation function for a two static droplet
source.  

\vspace*{0.3cm}
The case of $N=2$ presents an interesting example of the fluctuation
of the HBT correlation functions.  For the case of two static droplets
of $\sigma=2$ fm at an approximately touching configuration with a
separation of $2\sqrt{5} \sigma=8.84$ fm in the $x$ direction, Figs.\
(3$a$) give the correlation functions $C({\bf q})$ at $q_0=0$ as a
function for $q_z=0$, (3$b$) for $q_z=0.05$ GeV/c, and (3$c$) for
$q_z=0.10$ GeV/c.  All momenta in Fig.\ 3 are in units of GeV/c.  In
addition to the maximum at ${\bf q=0}$, there are prominent maxima and
minima at $q_x=2\pi/(R_1-R_2)$.  For this case, the correlation
function has minimum at $q_x \sim 0.70$ GeV/c and a maximum at $q_x
\sim 0.14$ GeV/c.  The correlation function decreases in magnitude as
$q_z$ increases.

In the case of two equal size droplets whose centers are moving with
equal and opposite velocities ${\bf v}_1=-{\bf v}_2=v {\bf e}_z$ along
the beam direction ($z$-axis), the correlation function $C(q)$ is
\begin{eqnarray}
\label{cq}
  C(q)&=&1 + \frac {e^{-(\sigma^2+\tau^2v^2)q_z^2 -\sigma^2{\bf q}_T^2-\tau^2q_0^2 }}{2}
\biggl \{ ~~~~~~ \cosh (2\tau^2q_0q_zv)
\nonumber\\
& & ~~~~~~~~~~~~~~
+~~\cos [{\bf q}\cdot({\bf R}_1-{\bf R}_2)-q_0(T_1-T_2)+q_z v(T_1-T_2)]
\biggr\}. 
\end{eqnarray}
In this case, the correlation function $C(q)$ is characterized by an
asymmetric distribution that is narrower in the $q_z$ direction than
in the transverse direction, in addition to a cosine-type modulation
along the direction of the initial separation of the two droplets.
The degree of the narrowing of the $q_z$ distribution width depends on
the of the emission time width $\tau$.  We show in Fig.\ 4 the
correlation function $C(q)$ at $q_0=0$ for two droplets with the same
geometrical dimensions as those in Fig. 3 but with $v_z \sim 1$ and
$\tau=\sigma$.  The general features of the correlation function
fluctuations are similar to the static case except that the
correlation function falls faster as a function of $q_z$, as one can
observe by comparing Figs. (3c) and (4c).

\vspace*{-0.3cm}
\begin{figure}[h]
\hspace*{1.4cm}
\psfig{file=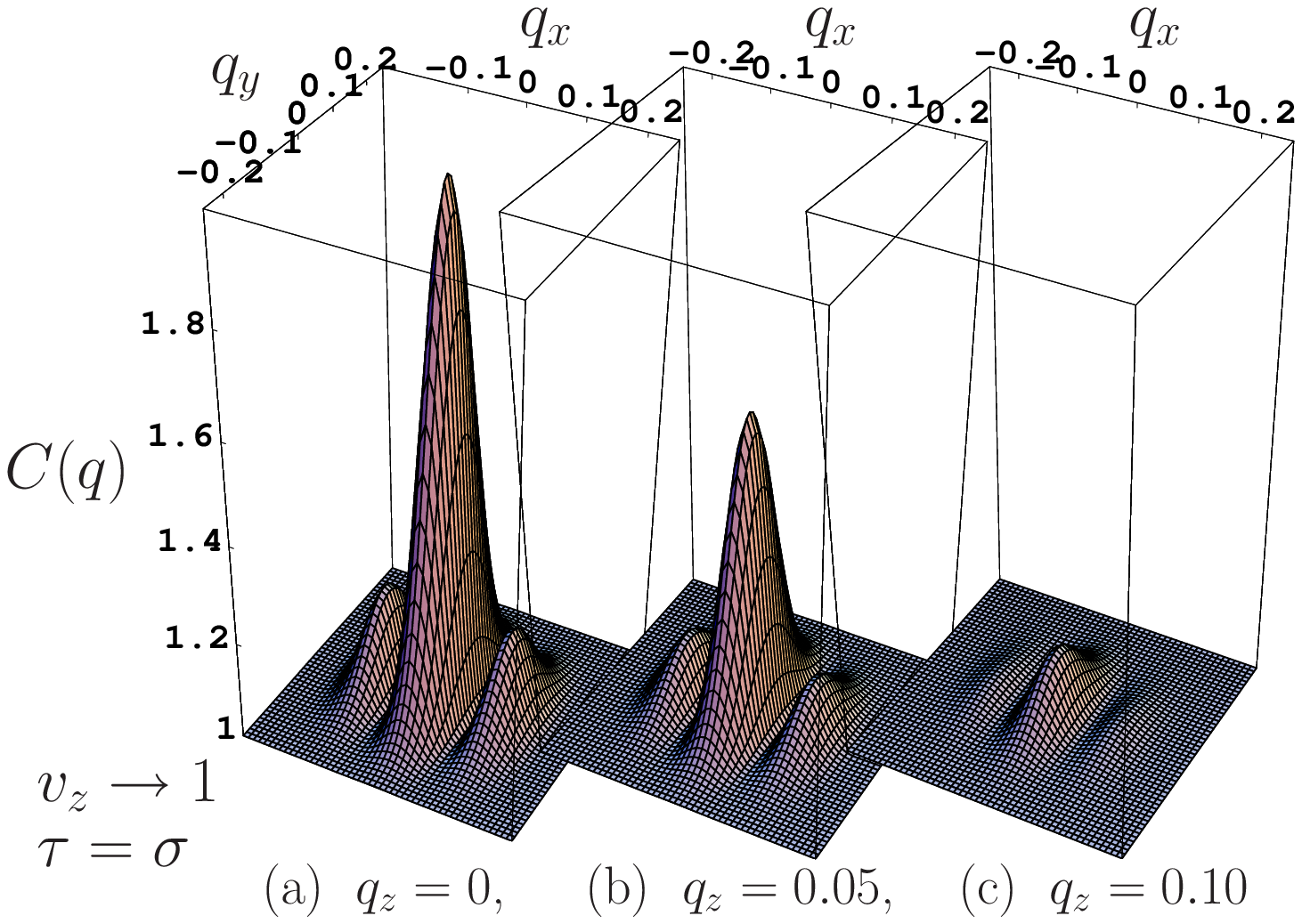,width=10cm}
\end{figure}

\vspace*{-0.4cm}
\hangafter=0
\hangindent=2.0cm
\noindent Fig. 4.  The correlation function for a two droplet source
with $v_z/c=1$ and $\tau=\sigma$.

\vspace*{0.3cm} In a single-event HBT measurement, one can locate the
reaction plane and designate the transverse axis in the reaction plane
to be the $x$-axis.  This $x$-axis can be considered as the axis
joining the centers of the two clusters initially.  Then the
correlation function will show prominent fluctuations in $q_x$, in
addition to the Gaussian distribution in $q_z$ and $q_T$, according to
Eq.\ (\ref{cq}).

If the statistics of single-event is insufficient, one can perform an
``aligned'' multi-event HBT analysis by referring to the transverse
axis in the reaction plane of each and every event as the $x$-axis.
Then, fluctuations in $q_x$ in different events are event-independent
and will be retained. A multi-event analysis with the properly aligned
transverse axis will exhibit large fluctuations as those shown in
Figs.\ 3 and 4 and indicated by Eq.\ (\ref{cq}).

On the other hand, one can use an ``unaligned'' multi-event HBT
analysis in which we do not align the transverse axis on the reaction
plane as the $x$-axis, ${\bf R}_1-{\bf R_2}$ from different events can
be considered randomly oriented in the transverse plane.  If the
emission time $T_i$ are the same, then the cosine term in Eq.\
(\ref{cq}) will be averaged to zero, and the correlation function
becomes
\begin{eqnarray}
\label{cq1}
  C(q)&=&1 + \frac {e^{-(\sigma^2+\tau^2v^2)q_z^2 -\sigma^2{\bf q}_T^2-\tau^2q_0^2 }}{2}
\cosh (2\tau^2q_0q_zv).
\end{eqnarray}
One can extract the radii $R_i({\rm HBT})$ from this type of
``non-aligned'' multi-event HBT analysis at $q_0\sim0$, corresponding
approximately to the inverse of the coefficient of $q_i^2$ in the
exponential factor in Eq.\ (\ref{cq1}), and one obtains
\begin{eqnarray}
R_x=R_y=R_{\rm side}=\sigma, ~~~~~{\rm and~~~~~} R_{\rm
long} \sim \sqrt{\sigma^2+\tau^2 v^2},
\end{eqnarray}
\begin{eqnarray}
R_{\rm long}/R_{\rm side}  \sim  \sqrt{\sigma^2+\tau^2 v^2}/\sigma.
\end{eqnarray}
This ratio is always greater than unity.  Thus, while the single-event
and the aligned multi-event analyses provide information on the
fluctuation of the correlation function, the unaligned multi-event
analysis will give the $R_{\rm long}/R_{\rm side}$ ratio to provide
useful information on the emission time and the relative velocity of
the two clusters.

\vspace*{-0.3cm}
\section{Discussions and Conclusions}

Density fluctuation occurs in high-energy heavy-ion collisions in many
ways.  It can arise from initial transverse density fluctuations
as a result of the underlying granular structure of the colliding
constituent nucleons.  It can come from the non-central character of
the collisions, leading to two or three clusters of pion emitting
clusters.  It can arise from the occurrence of first-order phase
transitions.  It can come from the surface tension effects which tend
to form clusters of strongly-interacting matter during its
evolution.  The surface tension effects also lead to the
multi-fragmentation of the produced matter into small clusters.  Some
of these density fluctuation sources are expected to be commonly
present and may be detected by single-event HBT interferometry.

In a nearly head-on collision at very high energies, the number of
identical pions is of the order of a few thousand.  The number of
observed identical pions $n_\pi$ is a small fraction of this number.
For example, the number of identical pions detected in the STAR
Collaboration in the most central Au-Au collisions at RHIC is of the
order of a few hundred\cite{Ada04}.  Although the number of pairs of
identical pions in the event varies as $n_\pi(n_\pi-1)/2$, only a
small fraction of these pairs have relative momenta small enough to
be useful in a HBT analysis.  Clearly, whether or not single-event
HBT measurements can be carried out remains to be tested.  

If the statistics of identical bosons turns out to be insufficient for
single-event analyses using the present detectors, there remains the
future prospect of performing single-event HBT analyses at RHIC with
detector upgrades or with heavy-ion collisions at LHC.  It will also
be of great interest to carry out few-event HBT analyses by using
pairs of pions in events with similar global characteristics.  Because
of the difference in initial conditions in different collision events,
different collisions will lead to different density fluctuations for
some degrees of freedom while retaining similar fluctuations for some
other degrees of freedom.  The use of few-event HBT analyses may still
retain those fluctuations from event-independent degrees of freedom.
This may be particularly useful in the examination of two or three
cluster configurations in non-central collisions by collecting events
of similar impact parameters and looking at pion pairs from these
events.  While the single-event and the aligned multi-event analyses
provide information on the fluctuation of the correlation function,
the multi-event analysis will give the $R_{\rm long}/R_{\rm side}$
ratio which will provide useful information on the emission time and
the relative velocity of the two clusters.

\vspace*{-0.3cm}
\section*{Acknowledgments}

This research was supported by the National Natural Science Foundation
of China under Contract No.10575024 and by the Division of Nuclear
Physics, US DOE, under Contract No. DE-AC05-00OR22725 managed by
UT-Battelle, LLC.

\end{document}